\documentclass[%
11pt,
]{iopart}

\usepackage{graphicx}% Include figure files
\usepackage{dcolumn}% Align table columns on decimal point
\usepackage{bm}% bold math
\usepackage{xcolor}
\usepackage{hyperref}  % TODO: any reason to avoid links?
\usepackage{enumitem}
\usepackage{printlen}
\usepackage{braket}
\usepackage[numbers]{natbib}
\newcommand{\Msun}{\ensuremath{\mathrm{M}_\odot}}
\newcommand{\Mc}{\ensuremath{\mathcal{M}}}
\begin{document}

%\preprint{LIGO-P1700020}

\title[GW150914 with future detectors]{How would GW150914 look with future gravitational wave detector networks?}

\author{Sebastian M. Gaebel and John Veitch}
\address{
School of Physics and Astronomy \& Birmingham Institute for Gravitational Wave Astronomy, University of Birmingham, Birmingham, B15 2TT, United Kingdom
 }
\ead{sgaebel@star.sr.bham.ac.uk}

\date{\today}

\begin{abstract}
The first detected gravitational wave signal, GW150914~\cite{gw150914_detection}, was produced by the coalescence of a stellar-mass binary black hole. Along with the subsequent detection of GW151226, GW170104 and the candidate event LVT151012, this gives us evidence for a population of black hole binaries with component masses in the tens of solar masses~\cite{o1_bbh}. As detector sensitivity improves, this type of source is expected to make a large contribution to the overall number of detections, but has received little attention compared to binary neutron star systems in studies of projected network performance.
We simulate the observation of a system like GW150914 with different proposed network configurations, and study the precision of parameter estimates, particularly source location, orientation and masses.
We find that the improvements to low frequency sensitivity that are expected with continued commissioning~\cite{observing_szenarios} will improve the precision of chirp mass estimates by an order of magnitude, whereas the improvements in sky location and orientation are driven by the expanded network configuration. This demonstrates that both sensitivity and number of detectors will be important factors in the scientific potential of second generation detector networks.
  
\end{abstract}

\maketitle

\def\convertto#1#2{\strip@pt\dimexpr #2*65536/\number\dimexpr 1#1}
\makeatother

\section{Introduction}
The first gravitational wave (GW) signal GW150914, from a black hole binary merger, was observed by the two Advanced LIGO (aLIGO, \cite{AdvLIGO}) detectors in Hanford and Livingston~\cite{gw150914_detection}. The masses of the two black holes were inferred to have masses $36.2^{+5.2}_{-3.8}$\,\Msun and $29.1^{+3.7}_{-4.4}$\,\Msun in their rest frame, forming a merger product of mass $62.3^{+3.7}_{-3.1}$\,\Msun~\cite{gw150914_pe,o1_bbh}.
The inferred source location was the target of follow-up observations by a range of instruments spanning the electromagnetic spectrum from radio to gamma rays~\cite{gw150914_emfollow}.
The sky localisation of this event was poorly constrained as it is largely determined by the difference in arrival time at the active detectors, and with only two operating aLIGO detectors the position was resolved to an annulus within a ring of constant time delay between the two sites~\cite{gw150914_pe}. However the Advanced Virgo (AdVirgo, \cite{AdVirgo}) detector is currently being commissioned and will join the network in July 2017 during the second observing run of Advanced LIGO, and KAGRA\,\cite{KAGRA} and LIGO-India\,\cite{LIGOIndia} to follow~\cite{observing_szenarios}. This raises the question as to how well those future networks can be expected to localize an event like GW150914, and how well its parameters could be measured with the upcoming second generation detector networks. The subsequent detection of GW151226~\cite{gw151226_detection}, GW170104~\cite{gw170104_detection} and the gravitational wave candidate LVT151012~\cite{gw150914_cbc_search,o1_bbh} provide evidence for a population of massive black hole binaries, which are likely to produce multiple further detections in the future~\cite{gw150914_rates,o1_bbh}.

Projections for future sensitivity improvements and network configurations are given in  \cite{observing_szenarios}, which also studies the sky location performance. However this study, in common with the majority of previous works~\cite{Berry:2014jja,Veitch:2012,Nissanke:2011,Singer:2014qca,Berry:2016}, considers only the binary neutron star case.
Expectations for localisation of generic systems were given in \cite{triangulation, WenChen:2010} using geometric arguments, which are a useful guide for qualitative interpretation of actual simulations in the $3+$ detector case. However, \cite{Grover:2014,Berry:2014jja} indicate quantitative differences between such arguments and full Bayesian parameter estimation results, and qualitative differences in the two-detector network from the availability of amplitude measurements.
Vitale \textit{et al}~\cite{Vitale_2GHeavy} studied the parameter estimation expectations for generic systems from a heavy BBH population which extends upward in mass above GW150914, while focusing mainly on mass and spin measurements. Most of the results are obtained using a network of one AdVirgo and two aLIGO detectors, although the five-detector network including LIGO-India and KAGRA was considered in an appendix but without comparing identical events. Essick \textit{et al}~\cite{Essick:2015} studied sky localisation for short transient signals, using generic burst algorithms, however these can be systematically different from sky localisation which uses a compact binary signal model~\cite{Vitale:2016jlv}.

In this article we address the question of localisation and parameter estimation for massive BH binaries from a different angle. Using GW150914 as a template, we perform a set of simulations based on an evolving network configuration, keeping the injected signals the same.
This allows us to study the improvements in parameter estimation and localisation systematically, using the initial Hanford-Livingston network as a reference, and studying the separate improvements produced by the expansion of the detector network and the general increase in sensitivity of these detectors.

For similar sources such as GW151226, LVT151012 or GW170104 we expect to see qualitatively similar behaviour, although for the lower mass systems the increased visibility of the inspiral portion of the signal will give a better overall constraint on the chirp mass. This is because the phase of the inspiral portion of the signal depends primarily on the chirp mass whereas the merger and ring-down portions depend more on the total mass (through the mass of the final BH), which can be seen in the comparisons in \cite{o1_bbh}. The lower mass signals also have a greater bandwidth in frequency, which should lead to a more precise localisation in general, although we expect the relative improvements from different network configurations to be similar.

We considered a variety of network configurations of GW detectors, based on the projections in~\cite{observing_szenarios}. We start with the sensitivity of the aLIGO Hanford and Livingston detectors in the first observing run (O1) and the eight engineering run (ER8b) which immediately preceded it, as a comparison point~\cite{gw150914_detector}, then add the Virgo detector with an initial noise curve as projected in \cite{observing_szenarios}. We compare this configuration to the network of Hanford, Livingston and Virgo at design sensitivity~\cite{AdvLIGO,AdVirgo}, and with a network expanded to include LIGO India~\cite{LIGOIndia}, KAGRA~\cite{KAGRA} and both.
Over the lifetime of the second generation instruments we expect the performance of the global network to improve parameter estimation in three important ways. The expansion of the global detector network will give better sky resolution and ability to better measure the signal polarisation (the Hanford and Livingston detectors are nearly co-aligned); the improvements at low frequency~\cite{observing_szenarios} will increase the observable duration of the signals and lead to more cycles of the inspiral part of the waveform being observable~\cite{parameters_from_waveform}; and finally the overall decrease in noise levels will greatly increase the signal-to-noise ratio (SNR) of the source. We investigate the effect of these improvements on sky localisation, mass measurement, and distance and inclination accuracy for a GW150914-like system.

\section{Method}
To compare the different network configurations we use a set of sixteen simulated signals and perform the full parameter estimation for each network set-up and each signal. The number of signals was chosen to balance computational cost and capturing the GW150914-like parameter space.
For both simulation and parameter estimation we used the reduced order model of the SEOBNRv2 waveform~\cite{seobnrv2rom,Taracchini:2013}, which models the inspiral, merger and ringdown parts of the GW signal and includes the effect of aligned spins on both component bodies.
The signal parameters were chosen to lie within the posterior distribution for the GW150914 event, so our simulations will appear to have the same relative amplitude in each detector as GW150914 did.
This allowed us to easily verify that the results appeared similar to GW150914 when using the Early Hanford-Livingston detector network.

We use a set of different network configurations which are designated by an identifier with three parts: The detector network, lower cut-off frequency, and noise spectrum. The detector network setup is a combination of aLIGO Hanford (H), aLIGO Livingston (L), AdVirgo (V), LIGO-India (I), and KAGRA (J). The noise spectrum is labelled as either ``Early'' , which indicates empirical ER8b/O1  spectra for H and L, and the projected early low curve for Virgo~\cite{observing_szenarios}, or ``Design'' for the expected sensitivities at final design specification~\cite{observing_szenarios,KAGRA}. The lower cut-off frequency is either $10~\mathrm{Hz}$ or $30~\mathrm{Hz}$. We selected twelve combinations of these to form the range of simulated network configurations. These configurations are shown along with results averaged over the 16 simulations in table~\ref{table:combined}.

We neglected $10~\mathrm{Hz}$ runs for early networks as they would yield little benefits due to the high noise levels at low frequency. KAGRA and LIGO India are still in construction phase and cannot be expected to start observing for some years, therefore the HLV detectors form the basis of the runs at design sensitivity. As the actual orientation of the future LIGO India detector was not available to us we assumed the arms to be aligned to North and East. Two-detector HL runs are included to represent the minimal possible configuration and to give results comparable to GW150914~\cite{gw150914_pe}.

\begin{figure}
\centering
\includegraphics{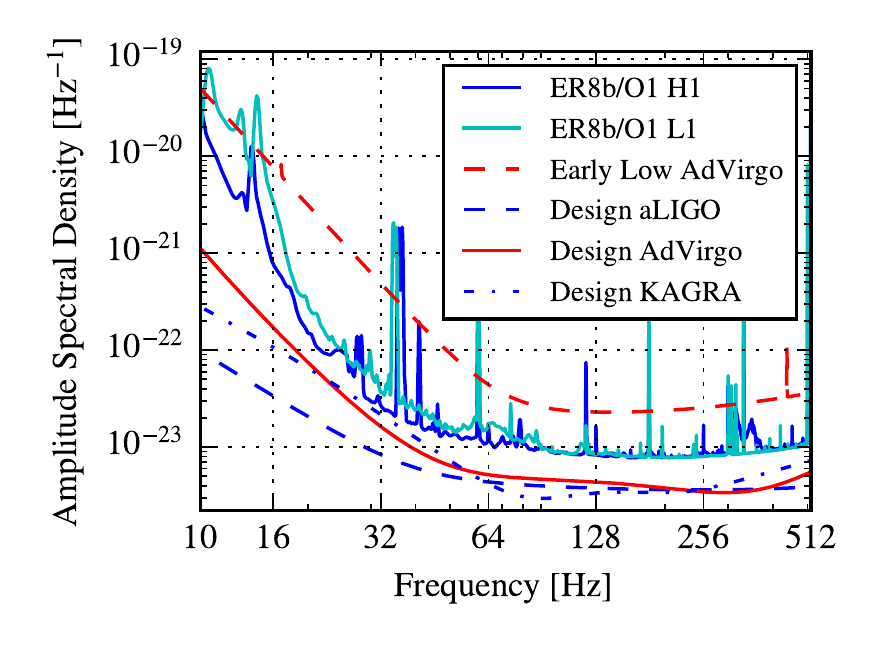}
\caption{The power spectral densities for the different sensitivities. The jagged curves for the ER8b/O1 curves are the median time averaged noise around GW150914~\cite{gw150914_noise}. All other curves are projected sensitivities~\cite{observing_szenarios}.}
\label{fig:psd}
\end{figure}

The power spectral densities for all sensitivity curves are given in figure~\ref{fig:psd}. The prior range of the component masses of the binary is $10\!-\!80~M_\odot$ so that it is wide enough to contain all possible simulations drawn from the GW150914 posterior and the posterior of those systems. Similarly, the duration of the data segment which was analysed was set to $8~\mathrm{s}$ for $30~\mathrm{Hz}$ runs and $160~\mathrm{s}$ for the $10~\mathrm{Hz}$ runs, based on the time spent in the analysed frequency band combined with a safety margin. We chose to set the noise realization to zero, meaning that the data contains only the simulated signal, so as to avoid noise perturbations affecting the comparisons. We assume no uncertainty in the phase and amplitude calibration of the detectors for our main results, although we additionally consider the effect of $10\%$ uncertainty in amplitude and $10^\circ$ in phase in the appendix. All parameter estimation was performed using LALInference~\cite{lalinference} in its nested sampling mode.

\section{Results}
Individual parameters are characterized by different features of the waveform, and therefore affected differently by the improvements in noise levels or network extensions. This reflects the distinction between intrinsic parameters, which are properties of the source itself, and extrinsic parameters, which are related to the relative positions and orientations of the source and the detectors. For this reason we present the results for different parameters in their individual sections, which also include the discussion of the results, and their comparison to the expected scalings with SNR that have been derived from analytic approximations in the literature.

Table~\ref{table:combined} contains an overview of the results for all discussed parameters with each run averaged over all simulations. The quantities used to measure the precision of the parameter estimation are the sizes of the 90\% credible interval (C.I.) or area (C.A.) for the chirp mass, distance, and sky area, and the value from the maximum likelihood sample ($\mathcal{L}_\mathrm{max}$) for the SNR.
% Additional material about table 1?

All figures in the section show the results for only one simulated signal to increase readability and to show qualitative behaviour. The combined results for all simulations are given in table~\ref{table:combined}

\begin{table}
\centering
% 150914_NOCAL
\begin{tabular}{|l|c|c c c c c|}
\hline
Network  & $f_\mathrm{min}$              & SNR           & Area [$\mathrm{deg}^2$] & $\Delta_\mathcal{M}$ [$\mathrm{M}_\odot$] & $\Delta_{D}$ [$\mathrm{Mpc}$] & $\Delta_{\theta_{\mathrm{JN}}} [\mathrm{rad}]$\\
 &      [Hz]         & at $\mathcal{L}_{\mathrm{max}}$ & 90\% C.A.               & 90\% C.I.                 & 90\% C.I.                 & 90\% C.I. \\
\hline
GW150914 (final)~\cite{o1_bbh} & 20 & $23.7$ & $230$ & $3.9$ & $340$ & --- \\
\hline
Early HL & 30     & $26.7 \pm 1.6$   & $ 183 \pm 34$   & $ 5.0 \pm 0.5$  & $306 \pm 30$ & ---\\  % 1.57 \pm 0.62
Early HLV & 30    & $27.6 \pm 1.7$   & $ 8.4 \pm 5.06$ & $ 4.9 \pm 0.5$  & $237 \pm 48$ & ---\\  % 1.33 \pm 0.65
Design HL & 10    & $75.9 \pm 4.8$   & $  31 \pm 7$    & $0.24 \pm 0.03$ & $215 \pm 39$ & ---\\  % 0.96 \pm 0.16
Design HL & 30    & $72.5 \pm 4.7$   & $  34 \pm 8$    & $ 2.1 \pm 0.4$  & $218 \pm 38$ & ---\\  % 0.99 \pm 0.20
Design HLV & 10   & $86.9 \pm 7.2$   & $0.57 \pm 0.20$ & $0.23 \pm 0.03$ & $179 \pm 42$ & $0.85 \pm 0.15$\\
Design HLV & 30   & $84.1 \pm 7.5$   & $0.54 \pm 0.18$ & $ 1.8 \pm 0.4$  & $179 \pm 42$ & $0.86 \pm 0.15$\\
Design HLVI & 10  & $99.9 \pm 15.6$  & $0.21 \pm 0.15$ & $0.20 \pm 0.03$ & $140 \pm 56$ & $0.65 \pm 0.19$\\
Design HLVI & 30  & $96.4 \pm 15.3$  & $0.19 \pm 0.15$ & $ 1.6 \pm 0.4$  & $140 \pm 55$ & $0.65 \pm 0.20$\\
Design HLVJ & 10  & $ 108 \pm  14$   & $0.14 \pm 0.10$ & $0.21 \pm 0.03$ & $ 95 \pm 53$ & $0.46 \pm 0.26$\\
Design HLVJ & 30  & $ 107 \pm  14$   & $0.13 \pm 0.10$ & $ 1.5 \pm 0.3$  & $ 98 \pm 51$ & $0.46 \pm 0.25$\\
Design HLVIJ & 10 & $ 119 \pm  20$   & $0.11 \pm 0.08$ & $0.19 \pm 0.03$ & $ 90 \pm 54$ & $0.44 \pm 0.25$\\
Design HLVIJ & 30 & $ 116 \pm  20$   & $0.10 \pm 0.08$ & $ 1.4 \pm 0.3$  & $ 92 \pm 53$ & $0.44 \pm 0.26$\\
\hline
\end{tabular}
\caption{This table shows the improvements in parameters estimation across network configurations. The columns contain the means of the corresponding values over all simulations with the standard deviation across the 16 values. The 90\% credible areas (90\% C.A.) were computed using the Skyarea Python module~\cite{skyarea} which estimates the minimum area enclosing 90\% of the source location probability distribution. The SNR values (SNR at $\mathcal{L}_\mathrm{max}$) are taken from the maximum likelihood sample of each posterior distribution, representing a best point estimate value. For chirp mass, distance and inclination we give the sizes of the 90\% credible intervals (90\% C.I.) $\Delta_\Mc$, $\Delta_D$ and $\Delta_{\theta_{JN}}$ respectively. $\Mc$ is the red-shifted mass, as observed in the detector frame. We omit the inclination angle values for the early networks and the 2-detector configurations which show a bi-modal degeneracy between $\theta_{JN}<\pi/2$ and $\theta_{JN}>\pi/2$. The 90\% contiguous credible interval for these bi-modal posteriors should not be directly compared to those for the uni-modal posteriors, when the breaking of the degeneracy produces such a qualitative change in the results. This can be seen in fig~\ref{fig:distance_and_theta} and is discussed in section~\ref{sec:dist_iota}.
GW150914 is included for comparison but needs to be used with care as those results assume a waveform low-frequency cut-off of $20$\,Hz, and allow a calibration uncertainty of $4.8\%$ and $8.2\%$ in amplitude and $3.2^\circ$ and $4.2^\circ$ in phase for Hanford and Livingston respectively~\cite{o1_bbh}.}
\label{table:combined}
\end{table}

\subsection{Signal to Noise Ratio}
\label{sec:snr}
We use the optimal signal-to-noise ratio, which we define as $\mathrm{SNR}=\sqrt{\braket{h | h} }$, as a metric for comparing the strength of a signal against the background. We define $\braket{\cdot|\cdot }$ as the noise-weighted inner product 
\[
	\braket{a|b} = 4 \Re \int_{f_\mathrm{min}}^{f_\mathrm{max}}\; \mathrm{d}\!f \frac{a^*\!(f)b(f)}{S(f)}
\]
with $h(f)$ and $S(f)$ being the waveform template and noise power spectral density respectively. $f_\mathrm{min}$ is the low frequency cut-off, which is chosen according to the noise properties so that the signal does not accumulate significant SNR below that value. $f_\mathrm{max}$ is chosen to be above the highest frequency contribution in the signal. This relation shows that both lowering the cut-off frequency $f_\mathrm{min}$ and decreasing the noise $S(f)$ improve the SNR by either increasing the interval over which the SNR can be accumulated, or increasing the integrand itself~\cite{o1_bbh,gw150914_cbc_search}. The amount to which these increase the value depends on the noise spectrum in the region of interest. Observing a signal in $N$ detectors is expected to increase the SNR by a factor of $\sqrt{N}$ relative to using a single detector only, and not taking sensitivity patterns into account, since SNR adds in quadrature for independent measurements.

\begin{figure}
\centering
\includegraphics{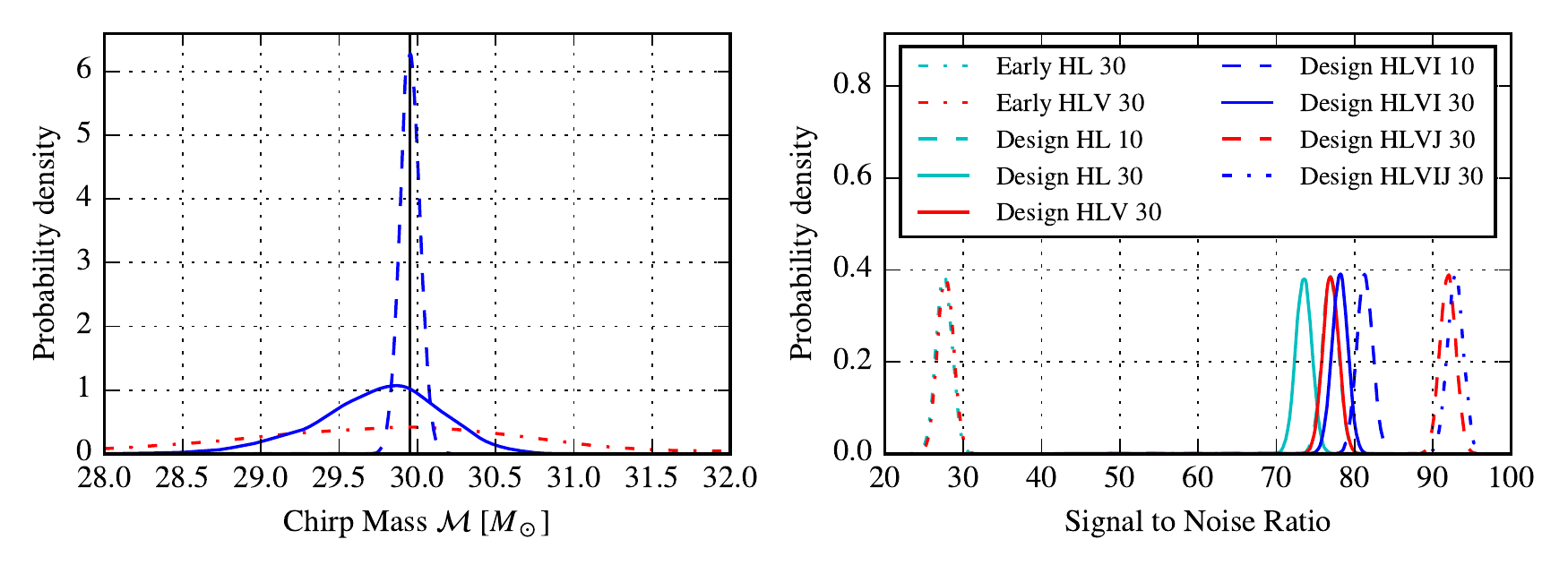}
\caption{The posterior distribution of the chirp mass (left) and SNR (right) for one individual signal as the network configuration changes (for overall differences in SNR see table~\ref{table:combined}. For the chirp mass only three runs are shown since the only factors which affect the distribution in a noticeable way are the switch to design sensitivity (red to solid blue), and the lowering of the lower cut-off frequency (solid blue to dashed blue). The main change in the SNR is caused by the switch to design sensitivity detectors, although additional detectors and a lower cut-off do have a noticeable impact.}
\label{fig:snr_mc}
\end{figure}

% Ratios below
Figure~\ref{fig:psd} shows that the noise levels start to rise quickly for frequencies lower than $\approx\!50~\mathrm{Hz}$ for all sensitivities. This explains why we see only minor differences in the SNR between $10~\mathrm{Hz}$ and $30~\mathrm{Hz}$ runs, which increases by a factor of only $1.01-1.05$. When increasing the detector sensitivity to the full design sensitivity however the SNR increases by a factor of $\approx\!2.7\!-\!3.0$. The gains from adding Virgo to the network in the low sensitivity case are minor, with a factor of $1.03$. This is, again, expected as the early low AdVirgo sensitivity is significantly less than that of the aLIGO detectors so it does not contribute much to the SNR. In the high sensitivity case the difference is noticeable with Virgo increasing the SNR by a factor of $\approx\!1.2$, bringing the total to $\approx\!84$ for the whole network. The fourth detector increases the combined SNR by a factor of $\approx\!1.2$ for LIGO India and $\approx\!1.3$ for KAGRA which suggests that KAGRA was in a more advantageous position for this event. Adding both LIGO India and KAGRA to the 3 detector setups brings the total SNR to $\approx\!116$, which is $\approx\!1.4$ times higher than the three detector value. The gains are roughly compatible with the expected values derived above, though we would not expect an exact match as the argument neglects differences in noise spectra and the impact of the antenna patterns for the different detectors. The measured SNRs for one simulation are shown in the right-hand panel figure~\ref{fig:snr_mc}, while the combined results for all simulations are available in table~\ref{table:combined}.

\subsection{Chirp Mass}
The chirp mass, defined as $
\mathcal{M} = (m_1 m_2)^{^{3\!}/_{\!5}}{(m_1 + m_2)^{^{-1\!}/_{\!5}}}
$ is the most important quantity in determining the frequency evolution for a GW from compact binaries.
Accordingly $\mathcal{M}$ can be measured precisely from the phase evolution of the waveform~\cite{parameters_from_waveform}, in contrast to extrinsic parameters such as the distance, which are measured from the signal amplitude as measured in multiple detectors.  Generally speaking, the measurement of the chirp mass improves due to the SNR according to the following relation for post-Newtonian inspiral signals~\cite{parameters_from_waveform}
\[
    \Delta (\mathrm{ln}~\mathcal{M}) \propto \; \mathrm{SNR}^{-1} \Mc^{^5/_3},
\]
which applies when the second order expansion of the posterior around the maximum is a good approximation, in the limit of high SNR~\cite{Vallisneri:2007ev}. We therefore expect the improved sensitivity to be helpful since it reduces the relative obfuscation of the waveform due to noise, increasing the SNR. Additionally, the sensitivity improvement at low frequencies, allowing for a reduced lower frequency limit for the observed signal, is expected to be beneficial as it enables us to detect additional cycles of the inspiral which contain information about the chirp mass. If each detector were equally sensitive the overall SNR would scale as $\sqrt{N}$, and therefore the measurement error would be expected to scale as $\frac{1}{\sqrt{N}}$ with $N$ being the number of detectors~\cite{parameters_from_waveform}.

The results for the full set of networks considered are shown in table~\ref{table:combined}. We report the detector-frame chirp mass measurements, which are affected by the red-shift of the source, but are the most easily comparable when looking at multiple systems which appear similar to the detectors. We find that with the ER8/O1 HL sensitivity the $90\%$ credible interval $\Delta_\mathcal{M}$ was a mean of $5.0$\,\Msun, which is slightly higher than the range of $3.9$\,\Msun\ reported in \cite{o1_bbh} for GW150914 using the SEOBNRv2 model, although this can be largely attributed to our use of $f_\mathrm{min}=30$\,Hz as opposed to 20 Hz. When using a lower cut-off of $20~\mathrm{Hz}$ we find a mean width of the 90\% credible interval of $3.1~\pm~0.3 \Msun$ which is slightly smaller than the GW150914 results, and to be expected since we assume perfect calibration and a zero-noise realisation.

When adding detectors we see minor gains, improving the chirp mass estimate by factors of $\approx\!1.02$ and $\approx\!1.04-1.15$ per detector added, for Early and Design sensitivity runs respectively. This is due to the relative sensitivity of the Advanced LIGO and Advanced Virgo instruments, such that the SNR increases less than the $\sqrt{N}$ formula implies. Improving the sensitivity proves much more rewarding, yielding an improvement factor of $\approx\!2.4-2.7$ when using the HL or HLV set-up at $30~\mathrm{Hz}$. The gains with a lowered frequency cut-off are even higher, improving the measurements by factor of $\approx\!7.3\!-\!8.7$. The left panel of figure~\ref{fig:snr_mc} shows a representative for each of the three distinct groups with nearly identical distributions. These groups are composed of the high cut-off, low sensitivity runs in the very wide case, the design sensitivity $30~\mathrm{Hz}$ runs for the intermediate peak, and the sharply peaked results from the two $10~\mathrm{Hz}$ runs.

\subsection{Sky localisation}

\begin{figure}
\centering
\includegraphics[width=1.0\textwidth]{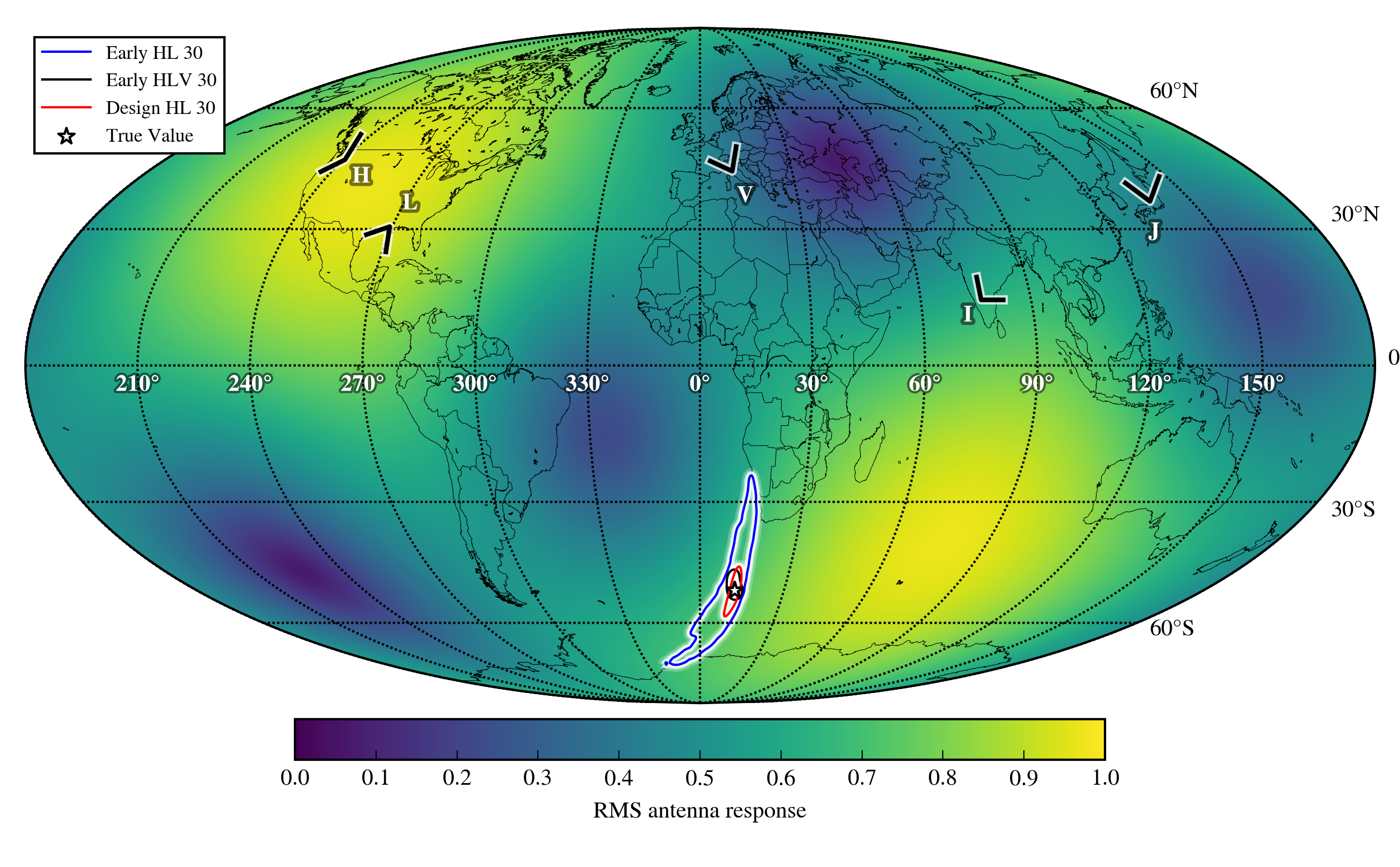}
\includegraphics[width=1.0\textwidth]{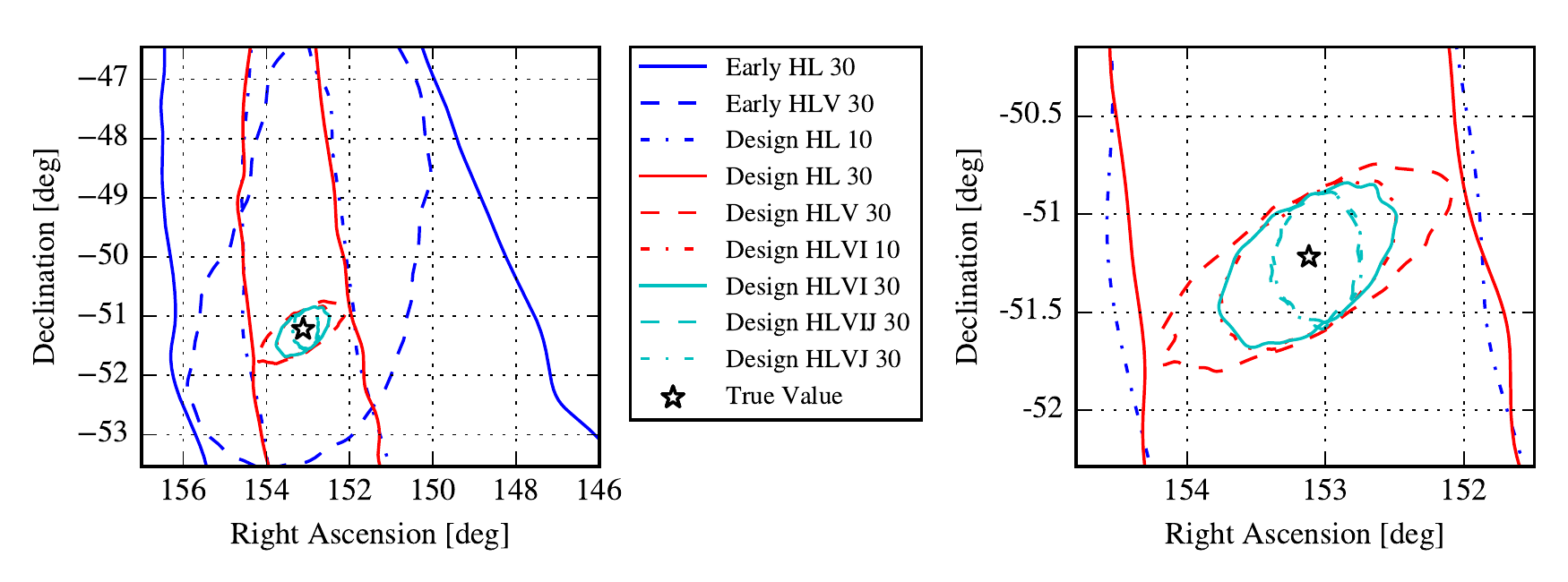}
\caption{The 90\% credible areas of one individual signal. The upper panel shows the global localisation over the root-mean-square sensitivity patterns of the Hanford and Livingston detectors added in quadrature. The two yellow areas mark the locations with the highest antenna response while signals from sources located in the dark blue regions are strongly suppressed. Only the largest areas are shown, as design sensitivity runs with 3+ detectors would be indistinguishable. The lower panels show the central region in greater detail and illustrate the differences between 3, 4, and 5-detector set-ups, which continuously shrink the area while remaining centred on the true location. Noteworthy is also that even at low sensitivity AdVirgo is able to improve the localisation massively and collapse the annulus into a region with the diameter comparable to the width of the 2-detector ring. The increased sensitivity has a greater impact in the 3-detector set-up as the improvements in AdVirgo are much larger.}
\label{fig:skyplots}
\end{figure}

The sky localisation is mainly determined by the timing measurements between the individual detectors~\cite{triangulation,observing_szenarios}. This means that there are two components to the measurement: the layout and synchronization of the detectors, and the measurement of the time delay using this external information. The main factor in how the layout of the detector network affects the sky localisation is in the distances between detectors. Larger baselines for the measurement of differences in arrival time translate into smaller relative errors, which then results in smaller uncertainties on the sky angles~\cite{triangulation}. The timing accuracy is inversely proportional to both SNR and the effective bandwidth~\cite{triangulation}. For small areas we can approximate the relevant section of the sphere as being flat, therefore the localisation is proportional to the square of the timing error, so we get:
\[
	\sigma_{\mathrm{area}} \propto \sigma_{\mathrm{RA}}\sigma_{\mathrm{Dec}} \propto \mathrm{SNR}^{-2}
\]
Even assuming perfect measurements, the nature of triangulation limits our ability to localize the source. Using only triangulation, with two detectors the source can be constrained to a circle, with three detectors to two points, and only the fourth detector allows us to narrow to location down to a single point. As the measurements are not perfect we do, however, still expect improvements from additional detectors beyond the fourth. Due to the fact that adding detectors does not only provide additional baselines for triangulation but also increases the SNR (see section~\ref{sec:snr}), we expect massive improvements in the sky localisation when detectors are added to the network. These gains should be the highest for the third detectors as it reduces the annulus to two single points, and to a lesser degree from the fourth detector which breaks the last degeneracy stemming from the symmetry under reflections on the plane of three detectors. Another advantage of an expanded detector network is rooted in the non-uniform antenna pattern of GW detectors which is shown in the upper panel of figure~\ref{fig:skyplots}. This causes detectors to have ``blind spots'' with low sensitivity, which can be compensated for by carefully choosing the position and orientation of other detectors. This helps to provide uniform sensitivity across the sky, and could increase the chances of making prompt electromagnetic follow-up observations of sources~\cite{Chen:2016luc}.

In addition to the timing triangulation, the relative amplitudes of the source in each detector, as determined by the angle-dependent antenna response functions, provides additional information about the position of the source which is naturally incorporated in our coherent analysis. This can break the ring-like or bimodal degeneracy in the two or three detector cases.

We observe that adding Advanced Virgo to the two Advanced LIGO detectors improves the localisation by factors of $\approx\!22$ and $\approx\!64$ for ER8b/O1 and design sensitivities respectively. Adding a fourth detector improves it by a factor of $\approx\!2.8$ when adding LIGO India or by $\approx\!4.0$ in the case of KAGRA over the HLV setup. The difference between these two possible 4 detector configurations is due to differences in sensitivity, as well as the antenna pattern. The full 5-detector configuration yields an area of $\approx\!0.1~\mathrm{deg}^2$ on average, which is smaller than the 3-detector result by a factor of $\approx\!5.3$. The areas range from tenths to hundreds of square degrees and are given in table~\ref{table:combined}.

\begin{figure}
\centering
\includegraphics{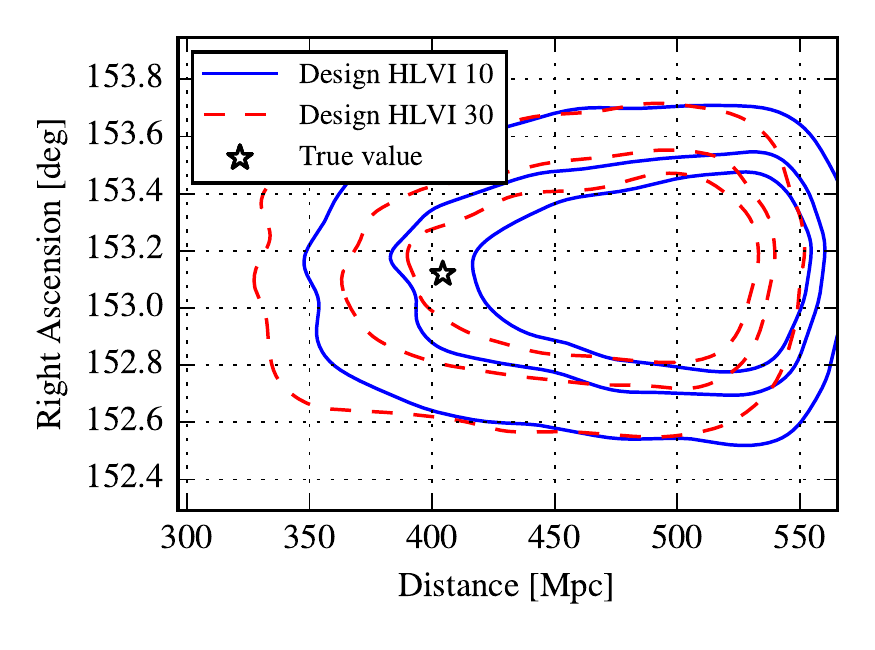}
\caption{The joint posterior distribution of the right ascension and luminosity distance shows that as the posterior moves to larger distances in the $10$\,Hz configuration the width of the marginal right ascension distribution increases. This behaviour is identical for the declination and thereby causes the localisation to worsen as low distance regions are excluded.}
\label{fig:distance_vs_ra}
\end{figure}

Unexpectedly, for the 3+ detector networks lowering the cut-off frequency did not improve the localisation despite the slight increase in SNR. We found that this is indirectly caused by a shift in the distance posterior, which causes the marginal distribution on sky angles to widen via the correlation shown in figure~\ref{fig:distance_vs_ra}, where the distribution for the angular parameters on the sky is larger at higher distances. While only the right ascension is shown this behaviour is identical for the declination.
The cause of the shift in distance seems to be that in the 10\,Hz runs the improvement in SNR and therefore amplitude uncertainty has translated slightly asymmetrically into a change in the distance posterior, influenced by the uniform-in-volume prior ($p(D)\propto D^2$) which tends to favour higher distances. The effect is small as can be seen from the relatively small change in $\Delta_D$, but it does seem to appear for all scenarios with 3+ detectors where the position is well constrained when comparing 10\,Hz and 30\,Hz cut-off frequencies.
In the two-detector case, the correlation between higher distances and higher areas is reversed, so that although the posterior is shifted a little the effect is to reduce the area by a factor $1.08$ not increase it. This effect seems to vary between different simulations in our set, depending on the precise geometry of the source and detectors so we do not believe this to be an important systematic trend.
The small difference can be seen in figure~\ref{fig:skyplots} by comparing the red dot-dashed and cyan solid lines in the lower panels.

\subsection{Distance and Inclination}
\label{sec:dist_iota}
\begin{figure}
\centering
\includegraphics{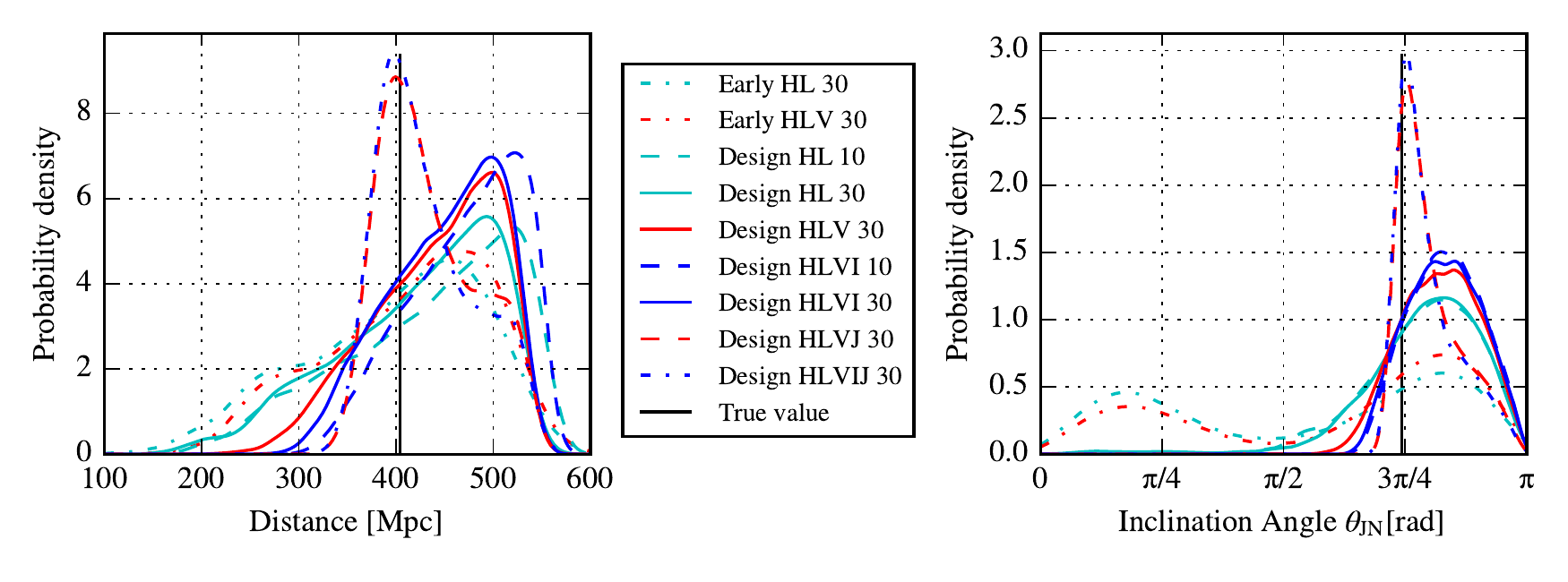}
\caption{The posterior distributions of the luminosity distance to the source and the inclination angle, for one simulated signal. The distance measurement covers a wide range of values, as the distance prior is uniform in volume and the distance is not very well measured and degenerate with other extrinsic parameters. The inclination angle is only weakly constrained at current sensitivities or with only two detectors. While the degeneracy between face-on and -off orientations can be broken with a third detector at higher sensitivities, the width of the peak decreases only from $\approx~\pi/4$ to $\approx~\pi/7$. Qualitative differences with the posterior peaking close to the maximum appear only once a fourth detector in an advantageous location is added.}
\label{fig:distance_and_theta}
\end{figure}

The inclination angle is the angle between the line of sight between the source and the observer $\vec{N}$, and the vector of the orbital angular momentum $\vec{L}$, which is aligned with the total angular momentum $\vec{J}$ in the aligned spins case considered here. 
It is a parameter which is typically weakly constrained by the GW observations, since it affects the relative amplitudes of the $+$ and $\times$ polarisations which are not individually resolvable by a single interferometer. Restricting to the dominant $l=m=2$ mode, the signal observed by the detector can be written as~\cite{Anderson:2000yy}
\[
	h(t) = \frac{1}{2} (1 + \cos^2(\theta_{JN})) F_+ A(t)\cos\Phi(t)  + \cos(\theta_{JN})F_\times A(t)\sin\Phi(t),
\]
with $\theta_{JN}$ being the inclination angle, $A(t)$, $\Phi(t)$ the amplitude and phase of the GW, and $F_+$, $F_\times$ the detector response functions for the $+$ and $\times$ polarisations, which depend on the relative position and polarisation of the source (see fig.~\ref{fig:skyplots}). As the two aLIGO detectors are nearly co-aligned they cannot on their own resolve both polarisations very well, leading to a degeneracy between left and right elliptically polarised waves, i.e. under the transformation $\theta_{JN}\mapsto \pi - \theta_{JN}$.
As the amplitude $A(t)$ is inversely proportional to the luminosity distance between source and observer~\cite{gw150914_pe,analytic_vt}, there is a further relationship between the inclination angle and the distance which allows edge-on nearby sources to appear similar to distance face-on (or face-off) sources. Together, these degeneracies produce the characteristic V-shaped posterior distributions as shown in e.g. fig.~2 of \cite{gw150914_pe}, and the bimodal $\theta_{JN}$ marginal distributions shown in the right panel of fig.~\ref{fig:distance_and_theta} for the HL networks and for the Early HLV network. 

While the inclination angle itself has little physical importance, the distance is important not only for the 3D source localisation, but also for the measurement of the masses in the source frame which needs to take the cosmological red-shift into account. This effect is already significant for GW150914 with a red-shift of only $\approx\!0.1$ and will only become more important for future detector networks, and especially third generation networks~\cite{Vitale:2016aso,Vitale:2016icu} as higher sensitivities greatly increase the number of observable sources at high distances.

Figure~\ref{fig:distance_and_theta} shows the posterior distribution for these two related parameters, with numeric values for the 90\% credible intervals $\Delta_D$ and $\Delta_{\theta_{JN}}$ available in table~\ref{table:combined}. The main feature is that both parameters are only weakly constrained for all network configurations.
We found that not only is the two-detector early HL network unable to break the degeneracy between $\theta_{JN}$ and $\pi-\theta_{JN}$ (left and right hand elliptically polarised waves), but the Early AdVirgo detector was not sensitive enough in comparison to Early aLIGO to do this either. With the HL network at design sensitivity most but not all of the signals were isolated to one of the $\theta_{JN}$ modes. As soon as three or more design sensitivity detectors are available the degeneracy was broken and the signal was isolated to only one of the lobes in $\theta_{JN}$ - the width of the 90\% credible intervals for these single-moded posterior distributions are shown in the rightmost column of table~\ref{table:combined}. This shows the benefit to having a global network of detectors that are able to measure the amplitude of both GW polarisations and therefore distinguish $\theta_{JN}$.
The addition of KAGRA makes a further qualitative difference to the results, improving the width of the 90\% credible interval by a factor of $\approx\!1.5-1.8$ and shifting the peak in figure~\ref{fig:distance_and_theta} to the true value, overcoming the prior which tends to prefer more distant, face-off orientations. In table~\ref{table:combined} this can be seen in the large differences between the 4-detector configurations, depending on whether they include KAGRA or LIGO India. This is the only qualitative change and illustrates the importance of detectors in various locations and orientations being available to break degeneracies and extract parameters. This behaviour is only observed without calibration uncertainty, the impact of introducing calibration uncertainty is discussed further in the appendix.

As there is an inverse relationship between distance and the measured signal amplitude, which is the quantity that is actually measured by the detectors, one might expect the fractional uncertainty on distance to scale as ${\Delta D}/{D} \propto {\mathrm{SNR}^{-1}}$, since the absolute uncertainty on amplitude is set by the noise level and the absolute value is proportionate to SNR~\cite{parameters_from_waveform}. However due to the correlations, the improved and extended detector network has a far lower effect on this set of parameters as compared to mass parameters. The size of the 90\% credible intervals for distance and inclination respectively decrease from $\approx\!306~\mathrm{Mpc}$ and $\approx\!\pi/4$ for the ER8/O1 2-detector network to $\approx\!90~\mathrm{Mpc}$ and $\approx\!\pi/7$ for the complete design sensitivity network. This under-performs in comparison to the improvement of $\sim4.5$ that one might expect from the $\mathrm{SNR}^{-1}$ scaling.

In a fashion similar to the slight worsening of the sky localisation, the size of the 90\% credible distance interval does not always decrease when switching to a $10~\mathrm{Hz}$ lower frequency cut-off. This is also caused by the small shift to higher distances, although the relative errors do decrease slightly as expected.

\section{Conclusion}

Although based around the GW150914 system, the results presented here give a good indication of the qualitative behaviour of parameter estimation for binary black hole systems as the global network of GW detectors continues to expand and improve in sensitivity. A less extensive subsequent analysis using the same procedure found that GW151226 shows a similar behaviour. There are minor differences caused by the lower mass which gives more importance to the inspiral over the merger and ring-down, as noted in~\cite{o1_bbh}. A large difference was observed in the ratios of sky areas caused by the initially poor localisation of the early HL network. This is highly dependent on the time delay between the two LIGO detectors: for systems which appear with a maximal time delay (~10\,ms) between the two sites the position is constrained to a small ring oriented near the projection of the vector between the sites onto the sky. On the other hand, if the time delay is near zero, as in the case of GW151226 (~1.1\,ms~\cite{gw151226_detection}), the ring has a large opening angle and therefore the projected area is much larger. Adding a third or more detectors to the network will mitigate this to a large degree~\cite{Fairhurst:2009tc}.

While the observed improvements in chirp mass were comparable to the approximate scaling relationship with SNR that may be derived from Fisher matrix calculations, they tend to under-perform slightly. This is expected since the detectors are not identical and the noise curves differ, especially at the low frequency end which is relevant for the chirp mass measurement. For distance and inclination, the behaviour is poorer due to the correlation and degeneracy shown in the posterior distribution, and so the scaling relationship based on the Fisher matrix expansion around a single maximum cannot hold, even for SNRs of 26 and above . Instead, the greatest effect comes from the expansion of the network and elimination of a large region of the sky, and the relative geometry of source and detectors. In general we expect the breaking of degeneracies to play an important role, but one that can vary significantly between different sky positions, as the relative detector responses change the amplitude of the signal in each detector.

With the combined improvements in sky localisation and distance measurement the volume to which future coalescence events will be constrained can be expected to decrease substantially as detectors are added and improved. As soon as a third detector joins the network the area which needs to be covered by electromagnetic observers decreases by factors of 20-60, which can be seen by comparing HL and HLV networks in Table~\ref{table:combined}. This will allow for a more complete coverage and greater depth to increase the chance of observing potential counterparts, or the (statistical) identification of a host galaxy~\cite{DelPozzo:2011yh,Singer:2016eax,Fan:2014}. Breaking the distance-inclination degeneracy will also aid the ability to perform cosmology with GW sources~\cite{Schutz:1986gp,Taylor:2011fs,DelPozzo:2015bna,Messenger:2011gi,Nissanke:2013fka}. For more detailed knowledge about the intrinsic properties of the sources themselves, the main driver is the improvement of the sensitivity. In case of the chirp mass the most important region is at low frequency where decreasing the cut-off from $30~\mathrm{Hz}$ to $10~\mathrm{Hz}$ can tighten the constraints by an order of magnitude.

In summary, although the field of gravitational wave astronomy as a true observational science has only just begun, the currently planned upgrades and expansions of the global network of detectors offer good observational prospects for heavy stellar mass binary black holes such as GW150914. Our work highlights the differing roles of (low-frequency) sensitivity and network geometry in aspects of constraining the source, indicating that a global network of comparable detectors will be necessary to achieve the best results for both mass estimates and source localisation.

\ack We thank Carl-Johan Haster, Simon Stevenson, Christopher Berry, and Alejandro Vigna-G\'{o}mez for useful discussions, and Salvatore Vitale for commenting on a draft of this manuscript. JV and SG were supported by STFC grants ST/K005014/1 and ST/M004090/1 respectively. We gratefully acknowledge UK Advanced LIGO computing resources supported by STFC grant ST/I006285/1.

\appendix

\section{Calibration Uncertainty}
In addition to the main network properties investigated above we replicated the analysis with a calibration uncertainty of $10\%$ in amplitude, and $10^{\circ}$ in phase, using the same interpolating spline model~\cite{SplineCalMarg-T1400682} as used in~\cite{gw150914_pe}, which is a conservative estimate of the uncertainty that may be expected for on-line calibration (and therefore relevant for initial parameter estimates)~\cite{Abbott:2016jsd}. At $10$\,Hz  we used only the HL and HLVI configurations due to large parameter spaces and consequent resource consumption.

The most significant differences appear in the extrinsic parameters. The sky localization worsens by a factor of $2.6-3.3$ for the early networks, and $7.5-10$ at design sensitivity. For distance and inclination angle the calibration improves the constraints by factors of $1.2-1.4$  for the HLVI configuration and $1.6$ to $1.8$ for networks including KAGRA. 
The observation of a sharp peak around the true value of both inclination and distance is a feature that starts to appear for 4+ design sensitivity networks. It is absent when using the $10\%/10^{\circ}$ calibration.
The chirp mass is affected to a much smaller degree. It worsens by factors of $1.1$ for early networks and $1.2-1.4$ at design sensitivity. The changes to SNR are on a level below $1\%$.

\begin{table}[h!bt]
\centering
% 150914_CAL
\begin{tabular}{|l|c|c c c c c|}
\hline
Network & $f_\mathrm{min}$                & SNR           & Area [$\mathrm{deg}^2$] & $\Delta_\mathcal{M}$ [$\mathrm{M}_\odot$] & $\Delta_{D}$ [$\mathrm{Mpc}$] & $\Delta_{\theta_{\mathrm{JN}}} [\mathrm{rad}]$\\
   &    [Hz]         & at $\mathcal{L}_{\mathrm{max}}$ & 90\% C.A.               & 90\% C.I.                 & 90\% C.I.                 & 90\% C.I. \\
\hline
GW150914 (initial)~\cite{gw150914_pe} & 20 & $25.1$ & $610$ & $4.4$ & $350$ & --- \\
\hline
Early HL & 30     & $26.7 \pm 1.8$  & $ 599 \pm 78$   & $ 5.4 \pm 0.4$  & $327 \pm 25$ & ---\\  % 2.6 \pm 0.1
Early HLV & 30    & $27.4 \pm 1.7$  & $  22 \pm 6$    & $ 5.3 \pm 0.4$  & $256 \pm 48$ & ---\\  % 2.4 \pm 0.2
Design HL & 10    & $75.7 \pm 4.7$  & $ 277 \pm 33$   & $0.33 \pm 0.04$ & $302 \pm 25$ & ---\\  % 1.7 \pm 0.6
Design HL & 30    & $72.3 \pm 4.6$  & $ 262 \pm 33$   & $ 2.6 \pm 0.5$  & $295 \pm 30$ & ---\\  % 1.9 \pm 0.5
Design HLV & 30   & $84.1 \pm 7.4$  & $ 5.5 \pm 0.9$  & $ 2.3 \pm 0.4$  & $219 \pm 43$ & ---\\  % 1.3 \pm 0.6
Design HLVI & 10  & $ 100 \pm 16$   & $ 1.7 \pm 0.5$  & $0.26 \pm 0.04$ & $191 \pm 38$ & $0.87 \pm 0.06$\\
Design HLVI & 30  & $96.3 \pm 14.8$ & $ 1.4 \pm 0.6$  & $ 2.0 \pm 0.4$  & $192 \pm 35$ & $0.88 \pm 0.05$\\
Design HLVJ & 30  & $ 107 \pm 14$   & $ 1.2 \pm 0.5$  & $ 2.0 \pm 0.4$  & $167 \pm 43$ & $0.76 \pm 0.18$\\
Design HLVIJ & 30 & $ 116 \pm 20$   & $0.85 \pm 0.27$ & $ 1.8 \pm 0.4$  & $163 \pm 43$ & $0.74 \pm 0.19$\\
\hline
\end{tabular}
\caption{This table is structurally identical to Table~\ref{table:combined} while presenting the results of the analysis using a calibration uncertainty of $10\%$ in amplitude and $10^{\circ}$ in phase.
It contains the means of the corresponding values over all signals with the standard deviation across the 16 values. The 90\% credible areas were computed using the Skyarea Python module~\cite{skyarea}. The SNR values are taken from the maximum likelihood sample of each posterior distribution. The values given for the chirp mass $\mathcal{M}$ and distance are the sizes of the 90\% credible interval. $\Mc$ is defined in the detector frame. We omit the inclination angle values for the two and three detector configurations, since the bi-modal posteriors are not well described by the 90\% credible interval (see fig~\ref{fig:distance_and_theta}).
In the first row we include the first GW150914 results, reported in \cite{gw150914_pe}, which includes comparable calibration uncertainty of $10\%$ in amplitude and $10^\circ$ in phase for both Hanford and Livingston. Note that this is different from the GW150914 (final) in Table~\ref{table:combined}.
}
\label{table:combined_cal}
\end{table}

Numeric values for all runs including calibration uncertainty are given in table~\ref{table:combined_cal}. We observe that, while the improvement of individual detectors and expansion of the network are important, improving the calibration is essential to obtaining the best possible results from the available detectors.

%\printbibliography
\bibliographystyle{iopart-num}
\bibliography{references}

\end{document}